# MMS Observations of a Compressed Current Sheet: Importance of the Ambipolar Electric Field

Ami M. DuBois[1], Chris Crabtree[1], Gurudas Ganguli[1], David M. Malaspina[2,3], William E. Amatucci[1]

[1]*U.S. Naval Research Laboratory, Plasma Physics Division, Washington, D.C. 20375-5346, USA*
[2]*Astrophysical and Planetary Sciences Department, University of Colorado, Boulder, CO 80303-7814, USA*
[3]*Laboratory for Atmospheric and Space Physics, University of Colorado, Boulder, CO, 80303-7814, USA*

Spacecraft data reveals a nonuniform ambipolar electric field transverse to the magnetic field in a thin magnetotail current sheet that leads to intense **E**×**B** velocity shear and non-gyrotropic particle distributions. The **E**×**B** drift far exceeds the diamagnetic drift and drives lower hybrid waves localized to the magnetic field reversal region, which is ideally suited for the anomalous dissipation necessary for reconnection. It also reveals substructures embedded in the current density, indicating the formation of a non-ideal current sheet.

Current sheets are important to space and laboratory plasmas [1–5], and particularly to the Earth's magnetotail where a consequential but overlooked transverse ambipolar electric field is self-consistently generated as the magnetotail is compressed by the solar wind. [6] This interaction of large and small-scale physics affects the plasma environment in the magnetosphere, also known as "space weather". In situ measurements show compression of the magnetotail that creates a thin current sheet of width comparable to the ion gyro-radius ($\rho_i$), occasionally with single [7–9] or double peaked [10–14] substructures in the current density. These are called non-ideal current sheets because their existence cannot be explained by the standard Harris equilibrium [15]. Intense lower hybrid wave activity and subsequent magnetic reconnection is also observed, which results in a plasma dipolarization front accelerating towards Earth, injecting energetic particles into the radiation belts. [16–21] It is, therefore, critical to understand the formation of non-ideal current sheets, especially the kinetic structures within, and the associated dynamics because their formation is thought to be important in the anomalous dissipation processes that initiate magnetic reconnection, which redistributes mass, energy, and momentum throughout the magnetosphere. [6,13,22–26]

Literature typically cites the lower hybrid drift instability (LHDI) as being associated with current sheets and magnetic reconnection because pressure gradients are observed. [17,27–31] However, theoretical [6,32,33] and laboratory [34–36] studies have shown that velocity-shear can intensify due to the ambipolar electric field generated by plasma compression and drive broadband turbulence peaking at the lower hybrid frequency. Shear-driven lower hybrid waves dominate over the LHDI when the shear frequency, defined as the spatial derivative of the $\vec{E} \times \vec{B}$ flow, exceeds the diamagnetic drift frequency. [6,37,38] Hence, identification of shear-driven lower hybrid waves in in-situ data is not only evidence of a non-ideal current sheet, but also emphasizes the importance of the ambipolar electric field [6], the scope of which has not been explored in previous magnetotail investigations [39–44]. As the scale size of the current sheet becomes comparable to $\rho_i$, the electric field intensifies making the velocity-shear strong, which can explain many observed features such as non-gyrotropic distributions [6,13,26,45,46], vortex structures [6,17,31,37,47], plasma heating and cooling [6,25,43], and the wide bandwidth of spectral signatures (broadband turbulence) [6,48]. There is evidence from laboratory experiments and ionospheric observations that velocity shear driven waves play important roles in ion heating [32,49–51], acceleration [32,37,52], transport [32,36,53,54], and other anomalous dissipation processes [32,37]. This letter presents the first evidence of shear-driven lower hybrid waves in a compressed current sheet that highlights the significance of transverse ambipolar electric fields in the structure and dynamics of thin current sheets.

The four Magnetospheric Multi-Scale (MMS) spacecraft [55] traversed a current sheet in the Earth's magnetotail on July 3, 2017 and crossed the null point at approximately 5:27:07.02 UTC. The measurements used to acquire the data are as follows: The magnetic field burst data are measured by the fluxgate magnetometer [56], ion and electron velocities, densities, and temperatures are obtained from the fast plasma investigation (FPI) [57] burst measurements, and electric field data are from the FIELDS [58] burst measurements. We find that rotating the vector data into a frame normal to the current sheet (i.e. LMN coordinates) suggests that this is a non-ideal current sheet undergoing compression. An ambipolar electric field forms due to the compression, giving rise to $\vec{E} \times \vec{B}$ shear flow in the same region as the observed lower hybrid fluctuations. The density gradient is small in this region, suggesting that the compression generated shear flow provides the energy for the lower hybrid fluctuations.

During this event, the MMS spacecraft are behind Earth on the dusk side of midnight far from the magnetopause. Using the spacecraft magnetic field measurements, minimum variance analysis (MVA) [59] is used to calculate the eigenvector corresponding to a direction normal ($\hat{N}$) to the current sheet. The angular difference between the averaged eigenvector and the eigenvectors calculated individually from each spacecraft ($\cos^{-1}(\hat{n}_{avg} \cdot \hat{n}_i)$) is small (~1°), which is a good indicator that the magnetic field gradient is much stronger in $\hat{N}$ compared to other directions. Additionally, a guide for judging the quality of the MVA is a large ratio of $\lambda_2/\lambda_3$ (eigenvalues corresponding to the $\hat{M}$ and $\hat{N}$ directions), ($\lambda_2/\lambda_3$=12) indicating a well-defined normal direction. [59] The vector data presented henceforth are rotated into LMN coordinates in order to better infer measurements with respect to the orientation of the current sheet such that $\hat{N}$ is normal to the current sheet, $\hat{L}$ is in the direction whose magnetic component reverses sign, and $\hat{M}$ is the direction of the guide magnetic field.

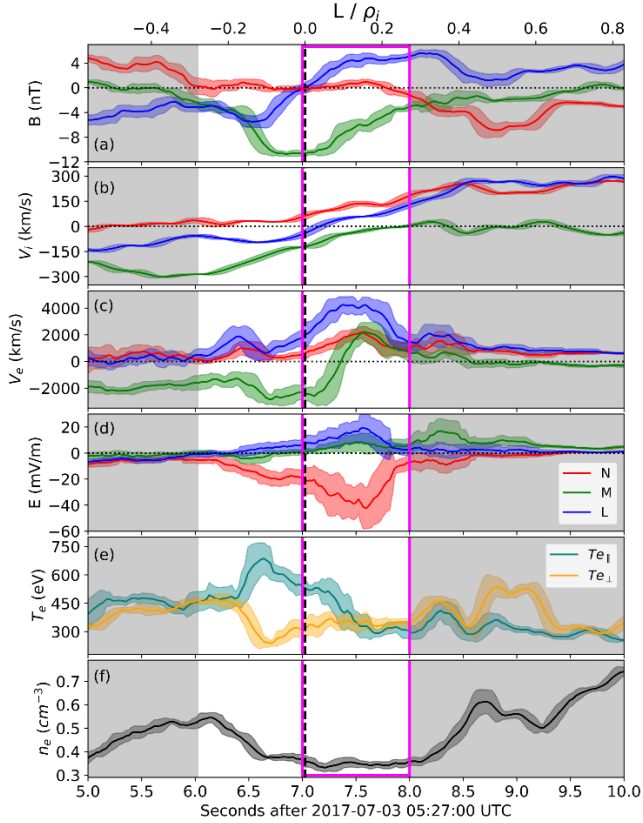

FIG. 1. Boxcar averaged MMS1 plasma parameters as a function of time (bottom axis) and distance normalized to a representative $\rho_i$ (top axis). (a) The magnetic field, (b) ion velocity, (c) electron velocity, and (d) the electric field rotated into LMN coordinates (blue=$\hat{L}$, green=$\hat{M}$, red=$\hat{N}$), (e) the parallel (teal) and perpendicular (orange) electron temperature, and (f) the electron density. The unshaded region highlights a ±1 second period around the $B_L$ reversal (vertical dashed line). The magenta box indicates the region during which electrostatic lower hybrid fluctuations are present.

A summary of the measured MMS1 data for a 5 second time span during the current sheet crossing is shown in Fig. 1. Data from each spacecraft are comparable with similar features, so only MMS1 data are shown. A boxcar averaging low pass filter routine is applied to the time series measurements to improve the signal to noise ratio and smooth the data to infer quasi-static profiles. The top axis displays time converted to distance and normalized to a representative $\rho_i$ (calculated from the average of the total magnetic field and ion temperature between 7 and 8 seconds) to show how features compare to ion-scale sizes. To convert time to distance, we use the fact that the current sheet is sweeping by in $\hat{N}$ much faster than the spacecraft are moving ($V_{MMS} = 1.47$ km/s). The magnetic field data from each spacecraft are shifted in time in order to minimize the difference between the time series. With the timing information, the spacecraft locations relative to each other, and the direction normal to the current sheet, the current sheet velocity can be estimated as $V_{cs} = 223.1 \pm 57.5$ km/s $\hat{N}$. The $V_{cs}$ is then multiplied by the timing of the measurements to calculate distance, and then normalized to $\rho_i$.

The magnetic field LMN components are shown in Fig. 1(a), where $\hat{L}$ is blue, $\hat{M}$ is green, and $\hat{N}$ is red. This shows that by rotating the vector data into a frame normal to the current sheet, $B_N$ is near-zero with a reversal in $B_L$ at 7.02 seconds (dashed vertical line), which indicates the current sheet crossing. The measurements are not symmetric about the $B_L$ reversal. There is a guide field ($B_M$) of approximately 10 nT at the time of $B_L$ reversal. Using the average total magnetic field between 7 and 8 seconds (magenta box), we calculate $\rho_i = 841$ km and $\rho_e = 6.9$ km (electron gyro-radius). The ion velocity [Fig. 1(b)] is significantly smaller than the electron velocity [Fig. 1(c)], which indicates that the ions experience negligible electric field due to gyro averaging as the electric field scale size is less than $\rho_i$ [6]. The total electron flow velocity, $V_{eL}$, has a scale size less than $\rho_i$. The electric field components are shown in Fig. 1(d), where $E_L$ and $E_M$ are both small compared to $E_N$ (the ambipolar electric field normal to the current sheet), which has a scale size less than $\rho_i$ and peaks at -45 mV/m. The electron temperatures ($T_e$) and density ($n_e$) are shown in Figs. 1(e) and (f). The perpendicular $T_e$ (orange) remains constant during the time window between 7 and 8 s, and the parallel $T_e$ (teal) decreases by ~300 eV. In the region of peak $E_N$, $n_e$ is nearly constant but exhibits a density gradient at approximately 6.5 and 8.5 s.

Electrostatic fluctuation spectra are calculated by taking a spectrogram of the $E_N$ burst measurement data prior to boxcar averaging. All panels in Fig. 2 show the electrostatic fluctuations from 2 to 20 Hz, where the color specifies the power spectral density such that black indicates the instrument noise level and yellow indicates large amplitude fluctuations. The lower hybrid frequency ($f_{LH} = \frac{1}{2\pi}\sqrt{\omega_{ci}\omega_{ce}}$) at the $B_L$ reversal is 7 Hz. The peak fluctuations occur at 7.5 s with a frequency of approximately 10 Hz, indicating these are lower hybrid waves. The boxcar averaged $E_N$ (solid, blue) and $n_e$ (dashed, orange) are overlaid on the fluctuation spectra [Fig. 2(a)] and shows that the electric field peaks during the largest amplitude fluctuations. The $n_e$ is nearly constant with practically no density gradient during the time of the fluctuations, and when there is a density gradient, no lower hybrid fluctuations are observed. As was shown in Fig. 1(a), the magnetic field in the region of peak $E_N$ is mostly $B_M$, meaning there is a large $\vec{E} \times \vec{B}$ velocity shear in $\hat{L}$ [Fig. 2(b), solid blue line], which peaks with the wave activity. The total electron flow in $\hat{L}$, $V_{eL}$ [Fig. 2(b), gray dot-dashed line], is in the same direction as $V_{E \times B}$, which indicates that the electrons are $\vec{E} \times \vec{B}$ drifting. Additionally, because $V_{eL}$ and $V_{E \times B}$ are close in magnitude, the diamagnetic drift velocity ($\vec{V}_{drift} = \nabla P_e \times \vec{B}/(n_e B^2)$) in the region of the wave localization is small compared to the $\vec{E} \times \vec{B}$ velocity. This is confirmed by the lack of the density gradient in this region, and indicates that the lower hybrid waves are driven by the sheared flow and not a density gradient.

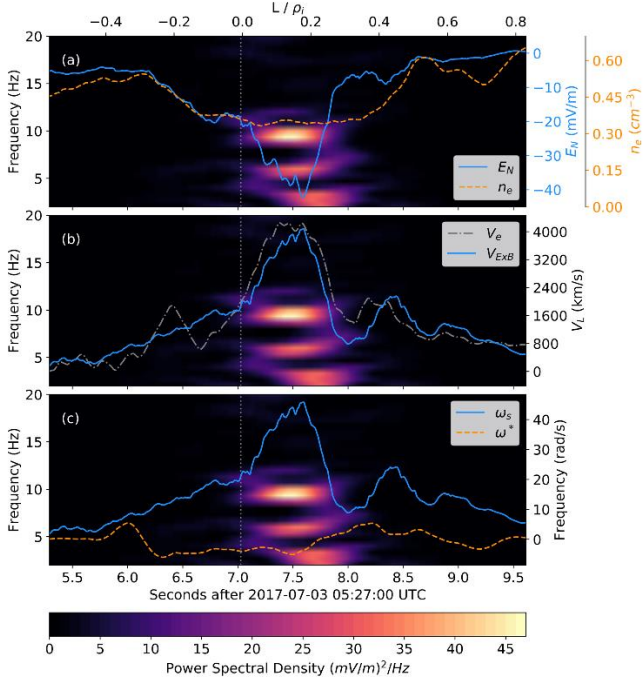

FIG. 2. Spectrogram showing power spectral density of electrostatic lower hybrid fluctuations as a function of time with (a) $E_N$ (solid, blue) and $n_e$ (dashed, orange) (b) total electron (gray, dot-dashed) and $\vec{E} \times \vec{B}$ shear (blue, solid) velocities in $\hat{L}$, and (c) the shear (blue, solid) and diamagnetic drift (dashed, orange) frequencies overlaid. The top axis shows distance normalized to $\rho_i$. The vertical dotted line indicates the $B_L$ reversal time.

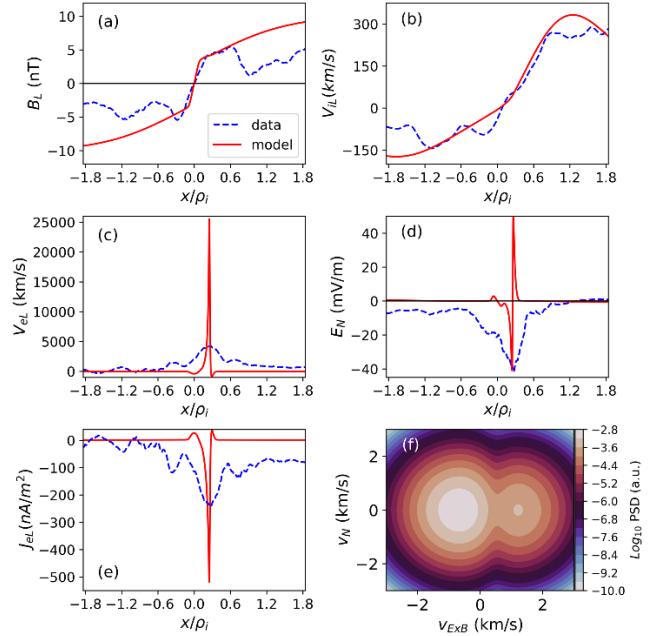

FIG. 3. Kinetic equilibrium model (solid, red) compared to MMS in situ data (dashed, blue) as a function of distance normalized to the ion gyro-radius for (a) magnetic field in $\hat{L}$, (b) ion velocity in $\hat{L}$, (c) electron velocity in $\hat{L}$ (d) ambipolar electric field in $\hat{N}$, (e) electron current density in $\hat{L}$. (f) The modeled non-gyrotropic electron distribution function that arises due to the ambipolar electric field, where the axes represent the electron flow in the direction of $E_N$ (vertical) and parallel to $V_{E \times B} \hat{L}$ (horizontal). The color bar represents the log of the phase space density in arbitrary units.

The origin of the observed features can be gleaned from our kinetic model [6,60], which is extended to include a guide field [61]. It constructs an exact solution to the Vlasov-Maxwell equations for species $\alpha$ by generalizing the Harris model to include an inhomogeneous guiding-center distribution,

$$f_{0\alpha}(x,v) = \frac{N_{0\alpha}}{(\pi v_{t\alpha}^2)^{3/2}} Q_\alpha(\Upsilon_\alpha, \zeta_\alpha) e^{-\frac{E_\alpha - U_\alpha p_y + \frac{1}{2} m_\alpha U_\alpha^2}{T_\alpha}}, \quad (1)$$

where $Q_\alpha(\Upsilon_\alpha, \zeta_\alpha) = G(\Upsilon_\alpha, L_{y\alpha}) H(\zeta_\alpha, L_{z\alpha})$, $G(\Upsilon_\alpha, L_{y\alpha}) = R_\alpha + S_\alpha \mathrm{Erf}((\Upsilon_\alpha - \Upsilon_0)/L_\alpha)$ is a non-negative, smooth ramp-like function (species dependent constants are defined in [6]), $H$ is a Gaussian function, and $\Upsilon_\alpha = [A_y(x)/B_0 + v_y/\Omega_\alpha]$ and $\zeta_\alpha = [A_z(x)/B_0 - B_v x + v_z/\Omega_\alpha]$ are the canonical momenta. These functions are chosen so that the moment integrals may be performed analytically. Then the quasi neutrality condition gives the electrostatic potential and Ampere's law gives the vector potential. This allows inhomogeneous structures in the moments (e.g. density, pressure, current, temperature, flows) to self-consistently develop in response to compression, which is represented by the scale sizes $L_{y\alpha}$ and $L_{z\alpha}$ of G and H, as in the case with no guide field described in [6]. This model was applied to the MMS data described above. Fig. 3 shows MMS1 data (dashed, blue) compared to the model (solid, red) for (a) $B_L$, (b) $V_{iL}$, (c) $V_{eL}$, (d) $E_N$, and (e) electron current density in $\hat{L}$. These calculations show that as a broad Harris-type current sheet undergoes global compression, an ambipolar electric field ($E_N$) self-consistently develops. This produces a sheared $\vec{E} \times \vec{B}$ velocity, enabling shear-driven waves to arise in thin current sheets. The presence of a guide magnetic field introduces sheared parallel flows, $V_{\parallel,L,M} = (\vec{E} \times \vec{B}) \cdot \hat{b}_{L,M}$, which can also drive waves [6]. Additionally, the bidirectional $V_{iL}$ flow profile is found to be a stationary Vlasov solution and therefore may not imply occurrence of magnetic reconnection, as claimed [17]. Given the initial conditions from the in situ measurements, the equilibrium model agrees with the general trends of the measurements. The observed profiles are smoother because the instability has relaxed the transverse gradients.

Our model also shows that as current sheets are compressed to scales less than $\rho_i$, substructures embedded in the current density can form, which cannot be explained by the standard Harris equilibrium. The current density ($J$) is calculated from MMS FPI data ($\vec{J} = en_i\vec{V}_i - en_e\vec{V}_e$), but because the $V_i \ll V_e$, the electron term dominates. Since $B_M > B_L$, the electron current density in $\hat{L}$, $J_{eL}$, is dominant. Fig. 3(e) shows the $J_{eL}$ data (dashed, blue) and model (solid, red). The large peak in $J_{eL}$ is localized to the region of the ambipolar electric field and the shear-driven lower hybrid waves. Both the data and the model suggest the formation of a non-ideal current sheet that has substructures contained within.

A distinguishing property of our equilibrium model is the formation of non-gyrotropic distribution functions [6], which have been seen in this event but were thought to be a wave heating effect [17]. Fig. 3(f) shows the electron distribution function from the model, where the vertical axis is the electron flow in the direction along the ambipolar electric field and the horizontal axis is the direction of $V_{E \times B}\hat{L}$. This shows that agyrotropy arises in the distribution function and is due to the ambipolar electric field that develops as a result of the compression.

Previous theoretical [6,32,37,60] and experimental studies [38] have shown that velocity shear can drive lower hybrid waves even in the presence of a pressure gradient as long as the shear frequency ($\omega_s \sim |\vec{V}_{E \times B}|/L_E$) is greater than the diamagnetic drift frequency ($\omega^* = -k_\perp \vec{V}_{drift}$). The shear scale length ($L_E$) can be estimated by taking the half-width at half-max of the boxcar averaged $E_N$. The average $L_E$ is estimated to be $89 \pm 22.9$ km such that $\rho_e / L_E < 1 < \rho_i / L_E$. Using the methods described in Norgren et al. [47], $k_\perp$ is calculated by first finding the time shift between $E_L$ measured from MMS1 and MMS2, which had the largest separation in $\hat{L}$. The wave phase speed is calculated using the distance in $\hat{L}$ between MMS1 and MMS2 and the time shift between the signals. The phase speed is then used to calculate the fluctuation wavelength, and $k_\perp = \frac{2\pi}{\lambda} = 0.027 \pm 0.01 \, km^{-1}$. Fig. 2(c) shows that $\omega_s$ (blue, solid) is an order of magnitude larger than $\omega^*$ (orange, dashed), indicating that the velocity shear is the dominating energy source for the observed waves over the pressure gradient. The Electron-Ion Hybrid (EIH) instability is driven by the free energy provided by sheared flows. [6,36,38,62] EIH waves are characterized by a $k_\perp \rho_e \ll 1$ and $k_\perp L_E \sim 1$. Observations from this event give $k_\perp \rho_e = 0.18 \pm 0.07$ and $k_\perp L_E = 2.4 \pm 1.2$ reinforcing the conclusion that the character of the lower hybrid fluctuations are consistent with EIH and not LHDI. Vortices, which have been reported earlier for this event [17], are also a natural consequence of the velocity shear that drives the EIH waves [37].

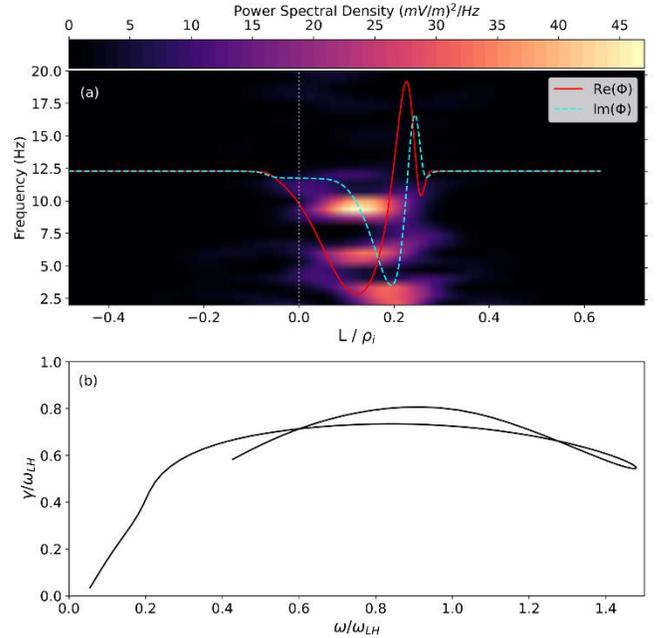

FIG. 4. (a) Spectrogram showing power spectral density of electrostatic lower hybrid fluctuations as a function of distance normalized to $\rho_i$ with the real (solid, red) and imaginary (dashed, blue) eigenmodes, ϕ, overlaid. The vertical dotted line indicates the $B_L$ reversal time. (b) The linear spectrum shows the growth rate (vertical axis) and frequency (horizontal axis) normalized to $\omega_{LH}$.

In order to compare this dataset with the EIH theory, we generalized the nonlocal theory of the EIH to include a guide field and magnetic field reversal. A model equilibrium electric field, consistent with the observed $E_N$, is used to drive sheared flows in $V_L$ and $V_M$. The eigenvalue problem is solved and a typical eigenmode, shown in Fig. 4(a), is localized around the strong $V_L$ flow near the center of the current sheet and is consistent with observations. Solution of the eigenmode condition as a function of $k_L$, $k_M$ provides a plot of growth rate versus frequency [Fig. 4(b)] normalized to $\omega_{LH}$ indicating a large domain of the shear driven instability that are ideally located in the center of the current sheet to contribute anomalous dissipation [37,53,54] for reconnection.

In summary, in-situ measurements show an ambipolar electric field develops in a thin current sheet in response to global compression and results in a strong sheared $\vec{E} \times \vec{B}$ velocity at the same time that a negligible density gradient is measured. Lower hybrid fluctuations are observed at the same time as the ambipolar electric field, and are determined to be shear-driven lower hybrid waves ($\omega_s \gg \omega^*$). Our equilibrium model captures the basic features of the in situ data and shows that a non-gyrotropic distribution function arises due to the ambipolar electric field and not necessarily due to wave heating. Magnetic field data rotated into LMN coordinates, the calculated current density, and the theoretical modeling indicate a highly compressed non-ideal current sheet with embedded substructures within. The identification of shear-driven lower hybrid waves is particularly significant because it indicates the importance of the ambipolar electric field in the structure and dynamics of thin current sheets, which has not been considered [39–44]. The results also show that velocity shear is an important source of free energy capable of driving lower hybrid waves in the magnetosphere, which are capable of anomalous processes that can trigger magnetic reconnection.

This work was supported by the Naval Research Laboratory Base Program. All MMS data used in this work are publically available and can be found online (https://lasp.colorado.edu/mms/sdc/public).